\documentclass[journal,onecolumn,12pt,draftcls,onecolumn]{IEEEtran}

\usepackage{epsfig,latexsym}
\usepackage{float}
\usepackage{indentfirst}
\usepackage{amsmath}
\usepackage{amssymb}
\usepackage{times}
\usepackage{subfigure}
\usepackage{psfrag}
\usepackage{cite}
\usepackage{color}
\definecolor{white}{rgb}{0.796,0.948,0.816}
\definecolor{black}{rgb}{0.00,0.00,0.00}

\def\diag{\textrm{diag}}

\newtheorem{theorem}{Theorem}
\newtheorem{corollary}{Corollary}
\newtheorem{lemma}{Lemma}

\begin{document}

\title{ {\color{black} On the Distribution of  GSVD}}

\vspace{-1em}
\author{Zhuo Chen,  Zhiguo Ding, \IEEEmembership{Senior Member, IEEE}
\thanks{
Zhuo Chen  is with Key Lab of Wireless-Optical Commun., Chinese Acad. of Sciences, Sch. Info Science \& Tech., Univ. Science \& Tech. China, Hefei, Anhui, 230027, P.R.China (E-mail: cz1992@mail.ustc.edu.cn, daixc@ustc.edu.cn).

Zhiguo Ding is with the School of Electrical
and Electronic Engineering, the University of Manchester, Manchester,
UK (E-mail: zhiguo.ding@manchester.ac.uk).

}}

\maketitle

\begin{abstract}
In this paper,  some new results on the  distribution of the generalized singular value decomposition (GSVD)
are presented.
\end{abstract}

\section{Some new results on GSVD}

In this section,
the GSVD of two Gaussian
matrices is defined first.
Then, the distribution of the squared generalized singular values
are presented.

\subsection{Definition of GSVD}
\label{GSVDdef}
Given two matrices $\mathbf{A} \in \mathbb{C}^{m \times n}$ and $\mathbf{C} \in \mathbb{C}^{q \times n}$,
whose entries are i.i.d. complex Gaussian random variables with
zero mean and unit variance.
Let us define  $k=\text{rank}((\mathbf{A}^H, \mathbf{C}^H)^H)=\min\{m+q,n\}$, $r=\text{rank}((\mathbf{A}^H, \mathbf{C}^H)^H)-\text{rank}(\mathbf{C})
=\min\{m+q,n\}-\min\{q,n\}$, and $s=\text{rank}(\mathbf{A})+\text{rank}(\mathbf{C})-\text{rank}((\mathbf{A}^H, \mathbf{C}^H)^H)=
\min\{m,n\}+\min\{q,n\}-\min\{m+q,n\}$.
Then, the GSVD of $\mathbf{A}$ and $\mathbf{C}$ can be expressed as follows \cite{cz2018}:
\begin{eqnarray}
\label{GSVDFORM}
\mathbf{U}\mathbf{A} \mathbf{Q}=\left(\mathbf{\Sigma}_\mathbf{A}, \mathbf{O} \right) \quad \text{and}  \quad \mathbf{V}\mathbf{C}\mathbf{Q}=\left(\mathbf{\Sigma}_\mathbf{C}, \mathbf{O} \right),
\end{eqnarray}
where $\mathbf{\Sigma}_\mathbf{A} \in \mathbb{C}^{m \times k} $ and $\mathbf{\Sigma}_\mathbf{C} \in \mathbb{C}^{q \times k}$ are two
nonnegative diagonal matrices, $\mathbf{U} \in \mathbb{C}^{m \times m}$ and $\mathbf{V} \in \mathbb{C}^{q \times q}$ are  two unitary matrices, and $\mathbf{Q} \in \mathbb{C}^{n \times n}$ can be expressed as in \eqref{expressQ}.

Moreover, $\mathbf{\Sigma}_\mathbf{A}$ and $\mathbf{\Sigma}_\mathbf{C}$ have the following form:
\begin{eqnarray}
\label{Diagform}
\mathbf{\Sigma}_\mathbf{A}= \left(\begin{array}{ccc}
\mathbf{I}_r&& \\
&\mathbf{S}_\mathbf{A}& \\
&&\mathbf{O}_\mathbf{A}
\end{array}\right)
\quad \text{and}  \quad \mathbf{\Sigma}_\mathbf{C}= \left(\begin{array}{ccc}
\mathbf{O}_\mathbf{C}&& \\
&\mathbf{S}_\mathbf{C}& \\
&&\mathbf{I}_{k-r-s}
\end{array}\right),
\end{eqnarray}
where  $\mathbf{S}_\mathbf{A}= \diag(\alpha_1,\cdots,\alpha_s)$
and $\mathbf{S}_\mathbf{C}=\diag(\beta_1,\cdots,\beta_s)$
are two $s \times s$ nonnegative
diagonal  matrices, satisfying
$\mathbf{S}_\mathbf{A}^2 + \mathbf{S}_\mathbf{C}^2 = \mathbf{I}_s$. Then, the squared generalized singular values
can be defined
as $w_i=\alpha_i^2/\beta_i^2$, $i \in \{1,\cdots,s\}$.

\subsection{Distribution of the squared generalized singular values}
\label{GSVDdistri}
To characterize the distribution of the squared generalized singular value $w_i$,
a relationship between $w_i$ and the eigenvalue of
a common matrix model is established first as in
the following theorem.

\begin{theorem}
\label{GSVDrelation}
Suppose that  $\mathbf{A} \in \mathbb{C}^{m \times n}$ and $\mathbf{C} \in \mathbb{C}^{q \times n}$ are two
Gaussian
matrices whose elements are i.i.d. complex Gaussian random variables
with
zero mean and unit variance
and their GSVD is defined as in \eqref{GSVDFORM}.
Without loss of generality, it is assumed that $q \geq m$.
Then, the distribution of their squared generalized singular values, $w_i, i \in \{1,\cdots,s\}$,
is identical to that of
the nonzero eigenvalues of $\mathbf{L}$, where
\begin{eqnarray}
\label{expressionL}
\mathbf{L}=\mathbf{X}^H \left(\mathbf{Y}\mathbf{Y}^H \right)^{-1}\mathbf{X}.
\end{eqnarray}
$\mathbf{X}\in \mathbb{C}^{m' \times p}$ and $\mathbf{Y}\in \mathbb{C}^{m' \times n'}$
are two independent Gaussian
matrices
whose elements are
are i.i.d. complex Gaussian random variables with
zero mean and unit variance.
Moreover, $m'$, $p$ and $n'$ can be expressed as follows:
\begin{eqnarray}
\label{weidu}
(m',p,n')=
\left\{ \begin{array}{ll}
(n,m,q)  &    q \geq n \\
(q,s,n) & q<n<(q+m)
\end{array} \right..
\end{eqnarray}
When $(q+m)\leq n$, $s=0$ and
$\mathbf{\Sigma}_\mathbf{A}$ and $\mathbf{\Sigma}_\mathbf{C}$ are deterministic.
\end{theorem}
\begin{IEEEproof}
See Appendix A.
\end{IEEEproof}

The distribution of the nonzero eigenvalues of $\mathbf{L}$
can be characterized as in
the following corollary.

\begin{corollary}
\label{Corro1}
Suppose that $\mathbf{L}=\mathbf{X}^H \left(\mathbf{Y}\mathbf{Y}^H \right)^{-1}\mathbf{X}$,
and $\mathbf{X}\in \mathbb{C}^{m' \times p}$ and $\mathbf{Y}\in \mathbb{C}^{m' \times n'}$
are two independent Gaussian
matrices
whose elements are
are i.i.d. complex Gaussian random variables with
zero mean and unit variance.
And, it is assumed that $m' \leq n'$.
Then, the joint probability density function (p.d.f.)  of the nonzero eigenvalues of $\mathbf{L}$
can be characterized as follows:
\begin{eqnarray}
\label{jointpdf}
f_{m',p,n'}(w_1,\cdots,w_{l})=  \mathbf{M}_{m',p,n'} \frac{   \prod_{i=1}^{l} w_i^{t_1}   }
{\prod_{i=1}^{l} (1+w_i)^{t_2}   }
\prod_{i<j}^{l} (w_i-w_j)^2,
\end{eqnarray}
where $l=\min\{p,m'\}$, $t_1=|m'-p|$, $t_2=p+n'$, and
$\mathbf{M}_{m',p,n'}$ can be expressed as follows:
\begin{eqnarray}
\label{Mexpress}
\mathbf{M}_{m',p,n'}=\left\{ \begin{array}{ll}
\frac{\pi^{m'(m'-1)} \widetilde{\Gamma}_{m'}(p+n')    }
{m'! \widetilde{\Gamma}_{m'}(p)  \widetilde{\Gamma}_{m'}(n') \widetilde{\Gamma}_{m'}(m') } &    p \geq m' \\
 \frac{\pi^{p(p-1)} \widetilde{\Gamma}_{p}(p+n')    }
{p! \widetilde{\Gamma}_{p}(m')  \widetilde{\Gamma}_{p}(p+n'-m') \widetilde{\Gamma}_{p}(p) } &  p < m'
\end{array} \right.,
\end{eqnarray}
where $m'!=1\times 2 \times \cdots m'$ and
$\widetilde{\Gamma}_{m'}(p)$ is the complex multivariate gamma function\cite{james1964distributions}.
\end{corollary}
\begin{IEEEproof}
Following steps similar to those in \cite[Appendix A]{2cz2018},
the distribution of the nonzero eigenvalues of $\mathbf{L}$ can be obtained.
\end{IEEEproof}

Moreover, the marginal p.d.f.
of $w_i$, $i \in \{1,\cdots,l\}$,
can be characterized as in the  following lemma.

\begin{lemma}
\label{lemma1}
The marginal p.d.f.
derived from \eqref{jointpdf}
can be expressed as follows:
\begin{eqnarray}
\label{pdayumpianmargin}
g_{m',p,n'}(w_l)=
  \mathbf{M}_{m',p,n'} g'_{l,t_1,t_2}(w_l) \quad \text{or} \quad
\frac{1}{w_l^2} \mathbf{M}_{m',p,n'} g'_{l,t'_1,t_2}(1/w_l),
\end{eqnarray}
where $t'_1=n'-m'$
and $g'_{l,t_1,t_2}(w_l)$ can be expressed as follows:
\begin{eqnarray}
\label{gpianeq}
g'_{l,t_1,t_2}(w_l)&=&\sum_{\sigma_1, \sigma_2 \in S_l} \mathbf{sign} (\sigma_1) \mathbf{sign} (\sigma_2)
\frac{w_l^{t_1+2l-\sigma_1(l)-\sigma_2(l)}}{(1+w_l)^{t_2}}
\\ \nonumber && \times
\prod_{i=1}^{l-1}B(t_1+2l-\sigma_1(i)-\sigma_2(i)+1,t_2-t_1-2l-1+\sigma_1(i)+\sigma_2(i)),
\end{eqnarray}
where $\sigma_1=\left(\sigma_1(1), \sigma_1(2) \cdots \sigma_1(l)\right)$, $\sigma_2=\left(\sigma_2(1), \sigma_2(2) \cdots \sigma_2(l)\right) \in S_l$ are permutations of length $l$, $\mathbf{sign} (*)$, $(*)\in\{\sigma_1,\sigma_2\}$, is $1$ if the permutation
is even and $-1$ if it
is odd,
and  $B(t_1+2l-\sigma_1(i)-\sigma_2(i)+1,t_2-t_1-2l-1+\sigma_1(i)+\sigma_2(i))$ is the Beta function  \cite{IntegralTable}.
\end{lemma}
\begin{IEEEproof}
See Appendix B.
\end{IEEEproof}

%and $_2F_1(-2,t_2-2l+i+j-t_1-1;t_2-2;1+w)$
%is the hypergeometric function

\subsection{Some properties about $\mathbf{Q}$}

As shown in \cite{cz2018},
the GSVD decomposition matrix $\mathbf{Q}$
is often used to construct the precoding matrix
at the transmitting end.
In this section,
some properties about $\mathbf{Q}$
are discussed.

First,
define
$\mathbf{B}=(\mathbf{A}^H, \mathbf{C}^H)^H$.
As shown in \cite[eq.(2.2)]{paige1981},
$\mathbf{Q}$ can be expressed as
\begin{eqnarray}
\label{expressQ}
\mathbf{Q}=\mathbf{Q}'\left(\begin{array}{cc}
\left(\mathbf{W}^H\mathbf{R}  \right)^{-1}&\\
& \mathbf{O}_{n-k}
\end{array}\right),
\end{eqnarray}
where $\mathbf{Q}'\in \mathbb{C}^{n \times n}$
and
$\mathbf{W} \in \mathbb{C}^{k \times k}$ are  two unitary matrices,
$\mathbf{R} \in \mathbb{C}^{k \times k}$
 is a nonnegative diagonal matrix and has the same singular values
as the nonzero singular values of $\mathbf{B}$.
Thus,
the power of
$\mathbf{Q}$ can be expressed as
\begin{eqnarray}
\label{powerQ}
\text{trace} \{ \mathbf{Q}\mathbf{Q}^H   \}&=&
\text{trace} \left\{  \mathbf{Q}'\left(\begin{array}{cc}
\mathbf{R}^{-1}\mathbf{W}&\\
& \mathbf{O}_{n-k}
\end{array}\right)
\left(\begin{array}{cc}
\mathbf{W}^H\mathbf{R}^{-H} &\\
& \mathbf{O}_{n-k}
\end{array}\right)  \mathbf{Q}'^H  \right\}
\\ \nonumber &=&
\text{trace} \left\{ \mathbf{Q}'^H \mathbf{Q}'\left(\begin{array}{cc}
\mathbf{R}^{-1}\mathbf{R}^{-H}&\\
& \mathbf{O}_{n-k}
\end{array}\right)
   \right\} \\ \nonumber &=&
\text{trace} \left\{  \left(\mathbf{R}^{H}\mathbf{R}\right)^{-1}
   \right\}
   \\ \nonumber &=&
\sum_i^k \lambda_{B,i}^{-1},
\end{eqnarray}
where $\lambda_{B,i}$, $i \in \{1,\cdots,k\}$,
are the nonzero eigenvalues
of $\mathbf{B}\mathbf{B}^H$.

Note that $\mathbf{A}$ and $\mathbf{C}$
are two independent Gaussian
matrices.
Then, it is easy to
know that
$\mathbf{B}\mathbf{B}^H$ is
a Wishart matrix.
Finally, directly from \cite[Lemma 2.10]{tulino2004foundations},
the following corollary can be derived.

\begin{corollary}
\label{Corro2}
Suppose that  $\mathbf{A} \in \mathbb{C}^{m \times n}$ and $\mathbf{C} \in \mathbb{C}^{q \times n}$ are two
Gaussian
matrices whose elements are i.i.d. complex Gaussian random variables
with
zero mean and unit variance
and their GSVD is defined as in \eqref{GSVDFORM}.
The average power of the GSVD decomposition matrix $\mathbf{Q}$
can be expressed as follows:
\begin{eqnarray}
\label{AverpowerQ}
\mathcal{E} \{ \text{trace} \{ \mathbf{Q}\mathbf{Q}^H  \} \}= \frac{\min \{m+q,n \}}{ |m+q-n|}.
\end{eqnarray}
\end{corollary}

\section*{Appendix A: Proof of Theorem \ref{GSVDrelation}}

\setcounter{subsection}{0}

\subsection*{1. The case when  $q \geq n$}

When $q \geq n$,  $k=\text{rank}((\mathbf{A}^H, \mathbf{C}^H)^H)=\min\{m+q,n\}=n$, $r=\text{rank}((\mathbf{A}^H, \mathbf{C}^H)^H)-\text{rank}(\mathbf{C})
=\min\{m+q,n\}-\min\{q,n\}=0$, and $s=\text{rank}(\mathbf{A})+\text{rank}(\mathbf{C})-\text{rank}((\mathbf{A}^H, \mathbf{C}^H)^H)=
\min\{m,n\}+\min\{q,n\}-\min\{m+q,n\}=\min\{m,n\}$. Thus, from \eqref{GSVDFORM}, the GSVD of $\mathbf{A}$ and $\mathbf{C}$ can be further expressed as follows:
\begin{eqnarray}
\label{GSVDFORMq>n}
\mathbf{U}\mathbf{A} \mathbf{Q}=\mathbf{\Sigma}_\mathbf{A}  \quad \text{and}  \quad \mathbf{V}\mathbf{C}\mathbf{Q}=\mathbf{\Sigma}_\mathbf{C} .
\end{eqnarray}
Moreover, when $k=n$, as shown in \eqref{expressQ}, $\mathbf{Q}$ is nonsingular and  it can be shown that
\begin{eqnarray}
\mathbf{A}^H\mathbf{A}= \mathbf{Q}^{-H} \mathbf{\Sigma}_\mathbf{A}^H \mathbf{\Sigma}_\mathbf{A} \mathbf{Q}^{-1}
\quad \text{and} \quad
\mathbf{C}^H\mathbf{C}= \mathbf{Q}^{-H} \mathbf{\Sigma}_\mathbf{C}^H \mathbf{\Sigma}_\mathbf{C} \mathbf{Q}^{-1}.
\end{eqnarray}
Furthermore,
from \eqref{Diagform},
it can be shown that
\begin{eqnarray}
\mathbf{\Sigma}_\mathbf{A}^H \mathbf{\Sigma}_\mathbf{A}=
\left(\begin{array}{cc}
\mathbf{S}_\mathbf{A}^H\mathbf{S}_\mathbf{A}&\\
& \mathbf{O}_{n-s}
\end{array}\right)
\quad \text{and}  \quad
\mathbf{\Sigma}_\mathbf{C}^H \mathbf{\Sigma}_\mathbf{C}=
\left(\begin{array}{cc}
\mathbf{S}_\mathbf{C}^H\mathbf{S}_\mathbf{C}&\\
& \mathbf{I}_{n-s}
\end{array}\right).
\end{eqnarray}
Thus, $\left(\mathbf{C}^H\mathbf{C}\right)^{-1}\mathbf{A}^H\mathbf{A}$ can be expressed as
\begin{eqnarray}
\left(\mathbf{C}^H\mathbf{C}\right)^{-1}\mathbf{A}^H\mathbf{A}= \mathbf{Q} \left(\begin{array}{cc}
\mathbf{S}_\mathbf{A}^H\mathbf{S}_\mathbf{A}\left[\mathbf{S}_\mathbf{C}^H\mathbf{S}_\mathbf{C}\right]^{-1}&\\
& \mathbf{O}_{n-s}
\end{array}\right) \mathbf{Q}^{-1}.
\end{eqnarray}
Recall that $\mathbf{S}_\mathbf{A}= \diag(\alpha_1,\cdots,\alpha_s)$
and $\mathbf{S}_\mathbf{C}=\diag(\beta_1,\cdots,\beta_s)$.
Then, it can be shown that
$\mathbf{S}_\mathbf{A}^H\mathbf{S}_\mathbf{A}\left[\mathbf{S}_\mathbf{C}^H\mathbf{S}_\mathbf{C}\right]^{-1}
=\diag(\alpha_1^2/\beta_1^2,\cdots,\alpha_s^2/\beta_s^2)
=\diag(w_1,\cdots,w_s)$.
Finally,
it is easy to see that
the distribution of  $w_i, i \in \{1,\cdots,s\}$,
is identical to that of
the nonzero eigenvalues of $\mathbf{L}$, where
\begin{eqnarray}
\mathbf{L}&=& \mathbf{A} \left(\mathbf{C}^H\mathbf{C} \right)^{-1}\mathbf{A}^H
\\ \nonumber
&=&\mathbf{X}^H \left(\mathbf{Y}\mathbf{Y}^H \right)^{-1}\mathbf{X}.
\end{eqnarray}
$\mathbf{X}\in \mathbb{C}^{m' \times p}=\mathbf{A}^H$ and $\mathbf{Y}\in \mathbb{C}^{m' \times n'}=\mathbf{C}^H$
are two independent Gaussian
matrices
whose elements are
are i.i.d. complex Gaussian random variables with
zero mean and unit variance.
Moreover, $(m',p,n')=(n,m,q)$.

\subsection*{2. The case when  $ q<n<(q+m)$}

When $q<n<(q+m)$, $k=\text{rank}((\mathbf{A}^H, \mathbf{C}^H)^H)=\min\{m+q,n\}=n$, $r=\text{rank}((\mathbf{A}^H, \mathbf{C}^H)^H)-\text{rank}(\mathbf{C})
=\min\{m+q,n\}-\min\{q,n\}=n-q$, and $s=\text{rank}(\mathbf{A})+\text{rank}(\mathbf{C})-\text{rank}((\mathbf{A}^H, \mathbf{C}^H)^H)=
\min\{m,n\}+\min\{q,n\}-\min\{m+q,n\}=m+q-n$.
As shown in \eqref{Diagform}, $w_1, \cdots, w_s$
are the squared generalized singular values
of $\mathbf{A}$ and $\mathbf{C}$,
and  $\alpha_i^2=\frac{w_i}{1+w_i}, i \in \{1,\cdots,s\}$.

On the other hand, define
$\mathbf{B}=(\mathbf{A}^H, \mathbf{C}^H)^H$
and the SVD of $\mathbf{B}$
as $\mathbf{B}=\mathbf{P}\mathbf{\Sigma}_\mathbf{B}\mathbf{R}$.
Note that
$\mathbf{P}\in \mathbb{C}^{(m+q) \times (m+q)}$
is a Haar matrix.
Divide
$\mathbf{P}$ into
the following four blocks:
\begin{eqnarray}
\label{Pblocks}
\mathbf{P}=\left(\begin{array}{cc}
\mathbf{P}_{11} & \mathbf{P}_{12}\\ \mathbf{P}_{21} & \mathbf{P}_{22}
\end{array}\right),
\end{eqnarray}
where $\mathbf{P}_{11} \in \mathbb{C}^{m \times n} $
and $\mathbf{P}_{22} \in \mathbb{C}^{q \times (m+q-n)} $.
From \cite[eq.(2.7)]{paige1981},
it is easy to see that
$\alpha_1^2, \cdots, \alpha_s^2$
equal the
non-one
eigenvalues of
$\mathbf{P}_{11}\mathbf{P}_{11}^H$.
Moreover, from
the fact that $\mathbf{P}$ is a Haar matrix,
it can be shown that
\begin{eqnarray}
\label{Punitary}
\mathbf{P}\mathbf{P}^H=\mathbf{P}^H\mathbf{P}=\mathbf{I}_{m+q}.
\end{eqnarray}
Thus, the following equations can be derived:
\begin{eqnarray}
\label{eqsP1}
\mathbf{P}_{11}\mathbf{P}_{11}^H+\mathbf{P}_{12}\mathbf{P}_{12}^H=\mathbf{I}_{m}
\quad \text{and} \quad
\mathbf{P}_{22}^H\mathbf{P}_{22}+\mathbf{P}_{12}^H\mathbf{P}_{12}=\mathbf{I}_{m+q-n}.
\end{eqnarray}
Note that $\mathbf{P}_{12}\mathbf{P}_{12}^H$
and $\mathbf{P}_{12}^H\mathbf{P}_{12}$
have the same
non-zero
eigenvalues.
Thus,
$\mathbf{P}_{11}\mathbf{P}_{11}^H$
and $\mathbf{P}_{22}^H\mathbf{P}_{22}$
have the same
non-one
eigenvalues.
Therefore,
$\alpha_1^2, \cdots, \alpha_s^2$
equal the
non-one
eigenvalues of
$\mathbf{P}_{22}^H\mathbf{P}_{22}$.
Define $\mathbf{P}'$ as
\begin{eqnarray}
\label{Ppian}
\mathbf{P}'=\left(\begin{array}{cc}
\mathbf{P}_{22}^H & \mathbf{P}_{12}^H\\ \mathbf{P}_{21}^H & \mathbf{P}_{11}^H
\end{array}\right).
\end{eqnarray}
It is easy to
see that
$\mathbf{P}'\mathbf{P}'^H=\mathbf{P}'^H\mathbf{P}'=\mathbf{I}_{m+q}$
and $\mathbf{P}'$ is also a Haar matrix.

Then,
from the above
discussions,
it can be concluded that
the distribution
of $\alpha_1, \cdots, \alpha_s$
is identical to
the distribution
of
the
non-one
singular valus
of the
$m \times n$
or $(m+q-n) \times q$
truncated
sub-matrix
of
a $(m+q) \times (m+q)$
Haar matrix.
Thus,
define
$\mathbf{A}' \in \mathbb{C}^{(m+q-n) \times q}$ and $\mathbf{C}' \in \mathbb{C}^{n \times q}$,
whose entries are i.i.d. complex Gaussian random variables with
zero mean and unit variance.
The distribution
of the squared  squared generalized singular values of $\mathbf{A}$ and $\mathbf{C}$,
is identical to that of
the squared  squared generalized singular values of $\mathbf{A}'$ and $\mathbf{C}'$.
Since $n>q$,
from Appendix A-1,
it can be known that
the distribution
of the squared  squared generalized singular values of $\mathbf{A}'$ and $\mathbf{C}'$,
is identical to that of
the  eigenvalues of
$\mathbf{L}
=\mathbf{X}^H \left(\mathbf{Y}\mathbf{Y}^H \right)^{-1}\mathbf{X}$,
where $\mathbf{X}\in \mathbb{C}^{q \times (m+q-n)}$ and $\mathbf{Y}\in \mathbb{C}^{q \times n}$
are two independent Gaussian
matrices
whose elements are
are i.i.d. complex Gaussian random variables with
zero mean and unit variance.

This completes the proof of  the theorem. \hspace{\fill}$\blacksquare$\newline

\section*{Appendix B: Proof of Lemma \ref{lemma1}}

\setcounter{subsection}{0}

First, the marginal p.d.f.
derived from \eqref{jointpdf}
can be expressed as follows:
\begin{eqnarray}
\label{marging1}
g_{m',p,n'}(w_l)&=&\int_0^{\infty}\cdots\int_0^{\infty}
\mathbf{M}_{m',p,n'} \frac{   \prod_{i=1}^{l} w_i^{t_1}   }
{\prod_{i=1}^{l} (1+w_i)^{t_2}   }
\prod_{i<j}^{l} (w_i-w_j)^2 dw_1\cdots dw_{l-1}
\\ \nonumber
&=& \mathbf{M}_{m',p,n'} \int_0^{\infty}\cdots\int_0^{\infty}
f'_{l,t_1,t_2}(w_1,\cdots,w_{l}) dw_1\cdots dw_{l-1}
\\ \nonumber
&=& \mathbf{M}_{m',p,n'} g'_{l,t_1,t_2}(w_l),
\end{eqnarray}
where
$f'_{l,t_1,t_2}(w_1,\cdots,w_{l}) dw_1\cdots dw_{l-1}
=\frac{   \prod_{i=1}^{l} w_i^{t_1}   }
{\prod_{i=1}^{l} (1+w_i)^{t_2}   }
\prod_{i<j}^{l} (w_i-w_j)^2$,
and
\begin{eqnarray}
\label{marging12}
 g'_{l,t_1,t_2}(w_l)&=&\int_0^{\infty}\cdots\int_0^{\infty}
f'_{l,t_1,t_2}(w_1,\cdots,w_{l}) dw_1\cdots dw_{l-1}
\\ \nonumber
&=&\int_0^{\infty}\cdots\int_0^{\infty}
\frac{   \prod_{i=1}^{l} w_i^{t_1}   }
{\prod_{i=1}^{l} (1+w_i)^{t_2}   }
\prod_{i<j}^{l} (w_i-w_j)^2 dw_1\cdots dw_{l-1}.
\end{eqnarray}
Note that $\prod_{i<j}^{l} (w_i-w_j)^2$ can be expressed as
\begin{eqnarray}
\label{W1}
 \prod_{i<j}^{l} (w_i-w_j)^2&=&|\mathbf{W}|^2
 \\ \nonumber &=&\left| \begin{array}{cccc}
w_1^{l-1} & w_1^{l-2} & \ldots & 1\\
w_2^{l-1} & \ldots  & \ldots & 1\\
\vdots & \ddots & \vdots& \vdots\\
w_{l}^{l-1} & \ldots  & \ldots & 1
\end{array} \right|^2.
\end{eqnarray}
Thus, $\prod_{i<j}^{l} (w_i-w_j)^2$ can be further expressed as
\begin{eqnarray}
\label{W2}
 \prod_{i<j}^{l} (w_i-w_j)^2=|\mathbf{W}|^2
=\sum_{\sigma_1, \sigma_2 \in S_l} \mathbf{sign} (\sigma_1) \mathbf{sign} (\sigma_2)
\prod_{i=1}^{l} w_i^{2l-\sigma_1(i)-\sigma_2(i)}.
\end{eqnarray}
Therefore, $g'_{l,t_1,t_2}(w_l)$ can be expressed as
\begin{eqnarray}
\label{marging133}
 g'_{l,t_1,t_2}(w_l)&=&
\sum_{\sigma_1, \sigma_2 \in S_l} \mathbf{sign} (\sigma_1) \mathbf{sign} (\sigma_2)
\int_0^{\infty}\cdots\int_0^{\infty}
\prod_{i=1}^{l} \frac{    w_i^{t_1+2l-\sigma_1(i)-\sigma_2(i)}   }
{ (1+w_i)^{t_2}   }
dw_1\cdots dw_{l-1}.
\\ \nonumber &=&
\sum_{\sigma_1, \sigma_2 \in S_l} \mathbf{sign} (\sigma_1) \mathbf{sign} (\sigma_2)
\frac{    w_l^{t_1+2l-\sigma_1(l)-\sigma_2(l)}   }
{ (1+w_l)^{t_2}   }
\prod_{i=1}^{l-1}\int_0^{\infty} \frac{    w_i^{t_1+2l-\sigma_1(i)-\sigma_2(i)}   }
{ (1+w_i)^{t_2}   }
dw_i.
\end{eqnarray}
Moreover, from Eq.(3.194.3) \cite{IntegralTable}, it can be shown that
$\int_0^{\infty} \frac{    w_i^{t_1+2l-\sigma_1(i)-\sigma_2(i)}   }
{ (1+w_i)^{t_2}   }
dw_i=B(t_1+2l-\sigma_1(i)-\sigma_2(i)+1,t_2-t_1-2l-1+\sigma_1(i)+\sigma_2(i)).$
Then,
$g'_{l,t_1,t_2}(w_l)$ can be expressed as
\begin{eqnarray}
g'_{l,t_1,t_2}(w_l)&=&\sum_{\sigma_1, \sigma_2 \in S_l} \mathbf{sign} (\sigma_1) \mathbf{sign} (\sigma_2)
\frac{w_l^{t_1+2l-\sigma_1(l)-\sigma_2(l)}}{(1+w_l)^{t_2}}
\\ \nonumber && \times
\prod_{i=1}^{l-1}B(t_1+2l-\sigma_1(i)-\sigma_2(i)+1,t_2-t_1-2l-1+\sigma_1(i)+\sigma_2(i)).
\end{eqnarray}

On the other hand, as shown in \eqref{marging12},
$g'_{l,t_1,t_2}(1/w_l)$
can be be expressed as
\begin{eqnarray}
\label{margingfan12}
g'_{l,t_1,t_2}(1/w_l)
&=&\frac{  w_l^{t_2-t_1}   }
{ (1+w_l)^{t_2}   }
\int_0^{\infty}\cdots\int_0^{\infty}
\frac{   \prod_{i-1}^{l-1} w_i^{t_1}   }
{\prod_{i=1}^{l-1} (1+w_i)^{t_2}   }
 \prod_{i=1}^{l-1} \left(w_i-\frac{1}{w_l}\right)^{2}
 \\ \nonumber && \times
\prod_{i<j}^{l-1} (w_i-w_j)^2 dw_1\cdots dw_{l-1}.
\end{eqnarray}
Moreover, define $w'_i=\frac{1}{w_i}$,  $i \in \{1,\cdots,l-1\}$. Then,
$g'_{l,t_1,t_2}(1/w_l)$
can be be further expressed as
\begin{eqnarray}
\label{margingfan1233}
g'_{l,t_1,t_2}(1/w_l)
&=&\frac{  w_l^{t_2-t_1}   }
{ (1+w_l)^{t_2}   }
\int_0^{\infty}\cdots\int_0^{\infty}
w_l^{-2(l-1)}
\frac{   \prod_{i-1}^{l-1} w_i'^{t_2-t_1-2l}   }{\prod_{i=1}^{l-1} (1+w'_i)^{t_2}   }
 \prod_{i=1}^{l-1} \left(w_i'-w_l\right)^{2}
 \\ \nonumber && \times
\prod_{i<j}^{l-1} (w_i'-w_j')^2 dw_1'\cdots dw_{l-1}'
 \\ \nonumber &=&
 w_l^2  \frac{  w_l^{t_2-t_1-2l}   }
{ (1+w_l)^{t_2}   }
\int_0^{\infty}\cdots\int_0^{\infty}
\frac{   \prod_{i-1}^{l-1} w_i'^{t_2-t_1-2l}   }{\prod_{i=1}^{l-1} (1+w'_i)^{t_2}   }
 \prod_{i=1}^{l-1} \left(w_i'-w_l\right)^{2}
 \\ \nonumber && \times
\prod_{i<j}^{l-1} (w_i'-w_j')^2 dw_1'\cdots dw_{l-1}'
 \\ \nonumber &=&
  w_l^2  g'_{l,t_1',t_2}(w_l),
\end{eqnarray}
where $t_1'=t_2-t_1-2l=n'-m'$.
Thus, it can be known that
$g_{m',p,n'}(w_l)
=\mathbf{M}_{m',p,n'} g'_{l,t_1,t_2}(w_l)
=\mathbf{M}_{m',p,n'} (1/w_l)^2 g'_{l,t_1',t_2}(1/w_l)$.

This completes the proof of  the lemma. \hspace{\fill}$\blacksquare$\newline

\bibliographystyle{IEEEtran}
\bibliography{IEEEfull,cz}
\end{document}